\documentclass[12pt, journal,onecolumn]{IEEEtran}

\begin{document}

\title{Carnot Efficiency of Publication}

\author{Abhisek~Ukil\\
Email: abhiukil@yahoo.com}

\maketitle


\section{Publication Efficiency}

Publications are increasingly (often wrongly) being used as research outputs. Furthermore, `impacts' of such publications are often (wrongly as well) measured using `citations'. Popular metric, like the `Hirsch index', or `h-index' is used as yardstick for judging the quality of individual researchers. However, there are many controversies around it, whether it is a good yardstick, which could be optimized, adultered, etc. `h-index' is also highly dependent on particular fields, which was pointed out by J. E. Hirsch himself in the second last paragraph in his paper \cite{hirsch}.

Nevertheless, if we consider citations as the output parameter, and papers as the input parameter for the publication system, it is worthwhile to estimate the efficiency of such process. The publication efficiency can be simply measured as
\begin{equation}
\label{eq:pub}
\eta_{pub}=\frac{h}{n},
\end{equation}

where, $h$ is the h-index and $n$ is the total number of papers of a researcher. Clearly, from eq. (\ref{eq:pub}), someone achieving particular h-index with less number of papers, would attain a higher publication efficiency. The maximum possible efficiency could be 1, i.e., when all the papers would contribute equally and optimally to the citations, hence the h-index.

Now, eq. (\ref{eq:pub}) can be simply modified as
\begin{equation}
\label{eq:pub2}
\eta_{pub}=\frac{h^2}{nh},
\end{equation}

In eq. (\ref{eq:pub2}), the denominator $nh$ indicates the total number of citations, i.e., $n$ papers getting $h$ citations each,
to contribute most optimally towards the h-index of $h$. Obviously, rarely someone's all paper would contribute optimally towards the h-index, i.e., usually $h^2 < nh$, in eq. (\ref{eq:pub2}).

\section{Relationship to Carnot Efficiency}
From eqs. (\ref{eq:pub} \& \ref{eq:pub2}), the most optimal publication system in terms of efficiency can be judged either by the ratio of h-index and total number of papers, or by the ratio of square of h-index and total citations. Between the two methods, the latter one can be computed more easily, as h-index and total citations can be easily obtained from publicly available data, e.g., Google Scholar or other databases. It can be conjectured that typically the
publication efficiency would attain a value around 30\% (with some variations), which seems to be rather field independent.

This can be equivalent to the `Carnot Efficiency' \cite{thermo} in thermodynamics, providing an upper limit on the efficiency that any classical thermodynamic engine can achieve during the conversion of heat into work. It is interesting to note that the efficiency of typical gasoline automobile engine is around 25\%, for large thermal electrical power plant is about 40\%, while Formula 1 motorsports cars peak around 45-50\% thermal efficiency \cite{thermo}.

\section{Usefulness of Publication Efficiency}
From the aforesaid discussions, it can be seen that the publication system, i.e., the process to convert paper into citations tend to follow the natural law like in thermodynamics. However, instead of maximizing the h-index or number of paper or citations, if one concentrates on maximizing the publication efficiency, it would naturally drive towards reduced quantity of high quality publication, attaining higher citations. This is probably increasingly needed in modern academic world, which is experiencing huge increase in number of publications, not necessarily of high quality. Adopting such measure would discourage publishing large number of incremental papers, which is becoming nuisance in certain fields, putting other fields in relative difficulty, especially the experimental ones.

\subsection{Example}
Let's imagine two cases, a senior researcher is settled with say h-index of 50 with 300 papers (typical numbers), while a young researcher starts with h-index of 10 with 20 papers. Which way the young researcher (assuming male subsequently) would go?

Current academic system will probably ask him to aim for publishing 280 more papers, which would kind of guarantee a respectable h-index. So, taking a step towards that, imagine in 5 years he publishes 100 more papers, making his h-index to 25. Other direction could be over next 5 years, he publishes 10 papers of high quality, making his h-index to 15. Is he doing good or bad? 

Traditionally, he would be bad in the second case. But in the first case, his efficiency is 25/(20+100)=21\%, while in second case, it is 15/(20+10)=50\% (very crude measure). And in the first case, with 100 papers in 5 years, he is more likely to produce less quality papers, than 10 papers in 5 years. However, most people would not risk taking the second path. Efficiency would at least motivate them to go probably for more quality, with some sort of mathematical support. Otherwise, no one will take such `risk' (should be `aim' rather), to write 50 landmark papers in life to eventually have an h-index of 50. 

Otherway, in case 1, if the researcher is not promoted until his efficiency is at least 30\%, he will naturally slow down, so that his h-index grows while not publishing too much non-sense paper. This way, one can at least discourage people from publishing large number of incremental papers for sometime, and possibly across all fields. 

Probably asymptotically one should target, that at the age of 50, one has 50 papers, all cited at least 50 times, with 100\% efficiency. This will be difficult, however, would possibly produce good works, maybe those 50 papers would be cited 500 times each (total citations 25000). If instead, one publishes 500 papers, each cited 50 times, it would still attain 25000 citations and 50 h-index. However, many will be diluted, and in the process, it would create a lot of confusions in the publication space, trying to beat the 1st law of thermodynamics like perpetual motion machines. Same h-index achieved with less papers is probably more efficient.


\end{document}